\documentclass[pra,floatfix,twocolumn,superscriptaddress]{revtex4}
\usepackage{times,graphicx,bbm,amsmath,amssymb,subfigure}
\usepackage{epsfig,color}
\usepackage{hyperref}

\def\Tr{\hbox{Tr}} 
\usepackage{pifont}

\newcommand{\ket}[1]{\vert#1\rangle}
\newcommand{\bra}[1]{\langle#1\vert}

\begin{document}

\title{Weak measurements and the joint estimation of phase and phase diffusion}

\author{Matteo Altorio} 
\affiliation{Dipartimento di Matematica e Fisica, Universit\`a degli Studi Roma Tre, Via della Vasca Navale 84, 00146, Rome, Italy}
\author{Marco G. Genoni} 
\affiliation{Department of Physics \& Astronomy, University College London, 
Gower Street, London WC1E 6BT, United Kingdom}
\author{Mihai D. Vidrighin}
\affiliation{QOLS, Blackett Laboratory, Imperial College London, London SW7 2BW, UK}
\affiliation{Clarendon Laboratory, University of Oxford, Parks Road, Oxford OX1 3PU, United Kingdom}
\author{Fabrizia Somma} 
\affiliation{Dipartimento di Scienze, Universit\`a degli Studi Roma Tre, Via della Vasca Navale 84, 00146, Rome, Italy}
\author{Marco Barbieri} 
\affiliation{Dipartimento di Scienze, Universit\`a degli Studi Roma Tre, Via della Vasca Navale 84, 00146, Rome, Italy}

\begin{abstract}
Weak measurements offer the possibility of tuning the information acquired on a system, hence the imposed disturbance. This suggests that it could be a useful tool for multi-parameter estimation, when two parameters can not be measured simultaneously at the quantum limit. Here we discuss their use for phase estimation in the presence of phase diffusion in the context of polarimetry, a scenario which is conveniently cast in terms of a two-level quantum system in many relevant cases. 
\end{abstract}

\maketitle

\section{Introduction}
Precision measurements often encounter intrinsic limitations imposed by quantum mechanics. Understanding these limitations and designing strategies for achieving the ultimate precision by means of quantum resources is the objective of quantum metrology~\cite{Giovannetti04}. 
The framework for single-parameter estimation is well established, both for Hamiltonian parameters, phases in particular~\cite{Holland93,Lee02,Higgins07,Brunn14}, and for relevant cases of dissipative parameters~\cite{Monras07,Adesso09}.

Physical processes, though, show both unitary and dissipative dynamics. A possible approach is to treat dissipation as a stationary process, which can be characterised with arbitrary precision, and consider phase estimation through the dissipative environment~\cite{Dorner09,Kacprowicz10,Datta11,Knysh11,Brivio10,Genoni11,Genoni12,Pinel13}. However, in the presence of non-stationary processes a more satisfactory approach consists in the joint estimation of the parameters linked to the unitary part and the evolution as well as to the dissipation. This requires tackling the problem with the formalism for multi-parameter estimation~\cite{Paris09}. Concerning the case above, when a first parameter is ascribed to a unitary and a second to the dissipation, this has been applied to the cases of phase and loss~\cite{Crowley14}, and phase with phase diffusion~\cite{Knysh14,Vidrighin14}.

This latter case exemplifies the subtleties of multi-parameter estimation. It has been recognised that there is no fundamental impediment to achieving the best possible precision for the phase shift $\phi$ and the magnitude of the phase diffusion $\delta$ simultaneously, and specific instances have been exemplified~\cite{Knysh14}. However, in practical cases, such as with coherent states, a trade-off appears, of which one can not dispose of by using standard quantum resources ~\cite{Vidrighin14}, such as $N00N$ states or Holland-Burnett states~\cite{Holland93}. This limitation appears to be intimately related with the geometric representation of these states, and only by carefully tailoring the quantum state one can circumvent these limitations in particular limits~\cite{Knysh14}. 

In this paper, we address this issue in polarimetry, in which the description is effectively carried out in terms of two-level systems, considering weak-measurement strategies. Their interest is in the possibility of repeated measurement of non-commuting observables~\cite{Lundeen,Howland}. There has been some debate on the scope of weak measurements in metrology, and on where they could offer some practical advantages~\cite{Dixon,Hofmann,Ferrie,Knee,Zhang,Howell}. Here we analyse how well weak measurements can be employed for multi-parameter estimation, and show under which conditions they saturate the trade-off for the estimation of $\phi$ and $\delta$ with the best possible precision.\\
The paper is structured as follows: in Sec. \ref{s:theory} we present the theory behind multi-parameter estimation 
of phase and phase-diffusion, and its relationship with weak measurements. In Sec. \ref{s:exp} we discuss an experimental realization of a weak measurement device based on an interferometric scheme, discussing its performances in the estimation of the two parameters of interest. We conclude the paper in Sec. \ref{s:conclusions} with remarks and possible outlooks.


\begin{figure}[b!]
\includegraphics[viewport = 50 350 550 550, clip, width=\columnwidth]{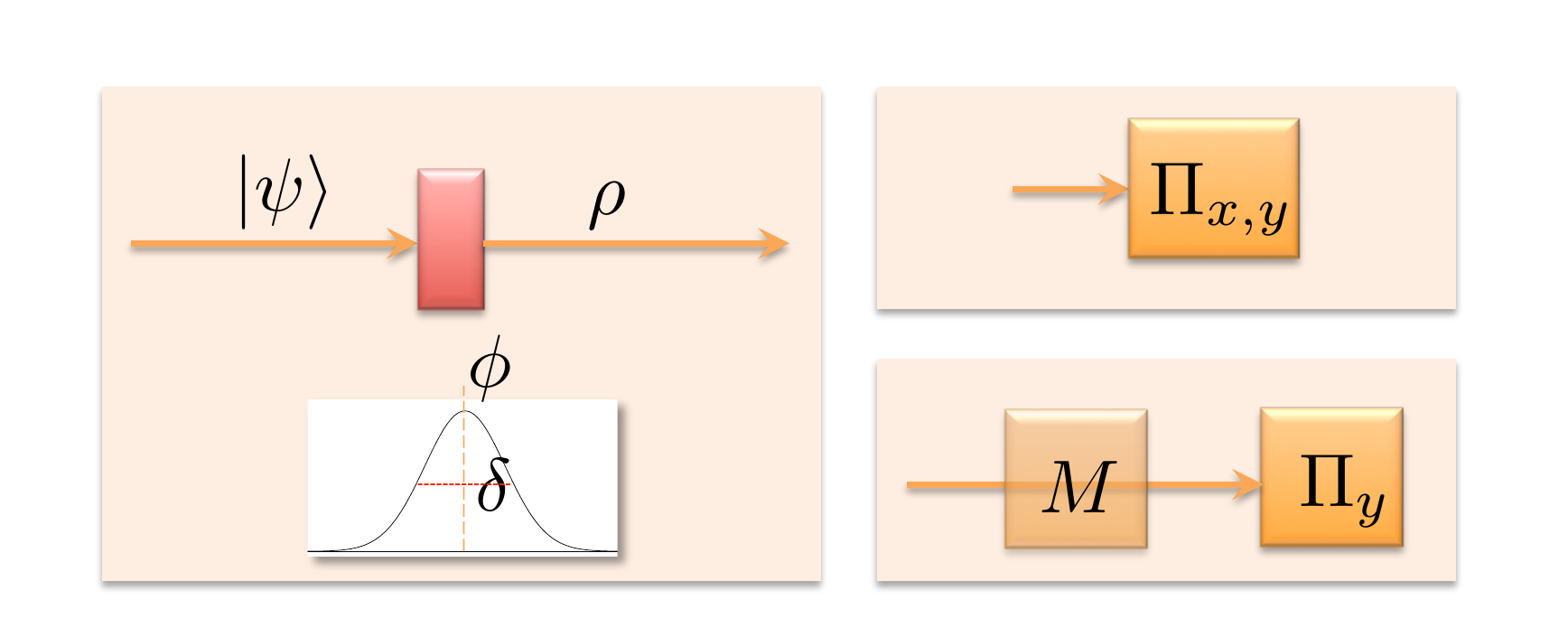}
\caption{(Color online) Schemes for multi-parameter metrology. A probe is prepared in the polarisation state ${\ket{\psi}=\frac{1}{\sqrt{2}}\left(\ket{y_+}+\ket{y_-}\right)}$, and sent in a sample imparting a random phase shift, whose distribution has a mean value $\phi$, and characteristic width $\delta$. Two possible schemes are considered: a four-outcome POVM $\Pi_{x,z}$, composed of projectors along the eigenvectors of $\sigma_x$ and $\sigma_z$, and known to saturate the bound \eqref{eq:1} as shown in \cite{Vidrighin14}; a weak-measurement $M$ along the $\sigma_z$ axis, followed by a projective measurement $\Pi_x$ along $\sigma_x$.}
\label{fig:1}
\end{figure}

\section{Theory} \label{s:theory}

Our ideal estimation experiment is illustrated in Fig.\ref{fig:1}. Polarised light passes through a sample, which imparts a birefringent phase; this is distributed around an average value $\phi$, with a distribution whose variance is $\delta$ (which will be refereed to as phase diffusion for brevity). The output density matrix, generated by the action of the channel on the optimal probe ${\ket{\psi}=\frac{1}{\sqrt{2}}\left(\ket{y_+}+\ket{y_-}\right)}$, can be written in the basis $\{ |y_\pm \rangle \}$ on which the phase shift occurs as~\cite{Paris09,Vidrighin14}:
\begin{equation}
\varrho=\frac{1}{2}\left(
\begin{array}{c c}
1 & e^{-i\phi-\delta^2}\\
e^{i\phi-\delta^2} & 1
\end{array}
\right).
\label{stato}
\end{equation}
\textcolor{black}{
The performance of an unbiased estimator for the two independent parameters $\phi$ and $\delta$ is quantified by the covariance matrix 
\begin{equation}
{\bold \Sigma} = \left(
\begin{array}{c c}
{\rm Var} (\phi) & {\rm Cov}(\phi,\delta) \\
 {\rm Cov}(\phi,\delta)  & {\rm Var} (\delta)
\end{array}
\right) .
\end{equation}
The covariance matrix, obtained via a number $M$ of repetitions of the experiment, is lower bounded as predicted by the {\em classical} and {\em quantum} Cram\'er-Rao bounds as follows
\begin{eqnarray}
M {\bf \Sigma} \geq {\bf F}^{-1} \geq {\bf H}^{-1} \label{eq:CRB}
\end{eqnarray}
where the matrices ${\bf F}$ and ${\bf H}$ are called respectively Fisher information and Quantum Fisher Information (QFI) matrices \cite{Paris09} (we report in the Appendix \ref{s:appendixLQE} general formulas for the two matrices). While the matrix ${\bf F}$ depends on the specific measurement (POVM) performed, the QFI matrix ${\bf H}$ is a property of the output state $\varrho$ only. In this particular instance, $\bold H$ is diagonal, hence we can find separate bounds for each variance independently: ${\rm Var}(\phi) \,H_{\phi\phi}\geq M$ and  ${\rm Var}(\delta)\,H_{\delta\delta}\geq M$ ~\cite{Knysh14,Vidrighin14}, where 
\begin{align}
H_{\phi\phi} &=e^{-2\delta^2} \nonumber \\
H_{\delta\delta} &=\frac{4\delta^2} {e^{2\delta^2}+1} \nonumber
\end{align}
are the diagonal elements of $\mathbf{H}$ associated with $\phi$ and $\delta$, {\it i.e.} the QFIs associated to each parameter~\cite{Vidrighin14}. 
}
Although $\mathbf{H}$ is diagonal, thus the parameters are in principle statistically independent, these can not be measured simultaneously at the quantum limit using a single two-level system, even when employing POVMs other than projectors and ensuring that no correlations between the two estimators arise~\cite{Vidrighin14}. This descends from the general result that in the multiparameter case the quantum Cram\'er-Rao bound is not tight, and may not be saturated~\cite{Paris09}. In fact, being  $\bold F$ the Fisher information matrix associated to any possible measurement operator on a two-level system, the following bound holds:
\begin{equation}
\label{eq:1}
\frac{F_{\phi\phi}}{H_{\phi\phi}}+\frac{F_{\delta\delta}}{H_{\delta\delta}}\leq 1 \:,
\end{equation}
\textcolor{black}{that, in turn, implies
\begin{equation}
\frac{F^\prime_{\phi\phi}}{H_{\phi\phi}}+\frac{F^\prime_{\delta\delta}}{H_{\delta\delta}}\leq 1 \:,
\label{eq:bound2}
\end{equation}
where we have introduced the effective Fisher information $F'_{\phi\phi}=1/(\bold F^{-1})_{\phi\phi}=F_{\phi\phi}-F^2_{\delta\phi}/F_{\delta\delta}$, and $F'_{\delta\delta}=1/(\bold F^{-1})_{\delta\delta}=F_{\delta\delta}-F^2_{\delta\phi}/F_{\phi\phi}$. As a consequence of Eq. (\ref{eq:CRB}), these quantities are in fact the ones bounding the variances for each estimator, given a specific measurement strategy, {\em i.e.} ${\rm Var}(\phi) F'_{\phi,\phi} \geq 1/M$ and ${\rm Var}(\delta) F'_{\delta,\delta}\geq 1/ M$~\cite{Paris09}. }{\color{black} By using the effective Fisher information, then, one explicitly takes into account the correlations between the parameters, thus collecting information about  one parameter, limits the information accessible on the other.}
Since in principle the right-hand side of Eqs. \eqref{eq:1} and \eqref{eq:bound2} can be as high as 2, this bound establishes that there exists no better strategy than using a fraction of the experimental runs to estimate the phase, and the rest for the phase diffusion, with a weighting giving the relative importance of the two parameters. A measurement saturating \eqref{eq:1} is described by a rank-1 POVM, measuring along the eigenvectors of $\sigma_x$ and $\sigma_z$. {\color{black} In this measurement strategy one can tune the relative weight of each parameter by tuning the probability of performing the $\sigma_x$ or the $\sigma_z$ measurement.} We observe that this bound is related to a similar relation established in~\cite{MassarGill} for state estimation.

\begin{figure}[t]
\includegraphics[width=0.9\columnwidth]{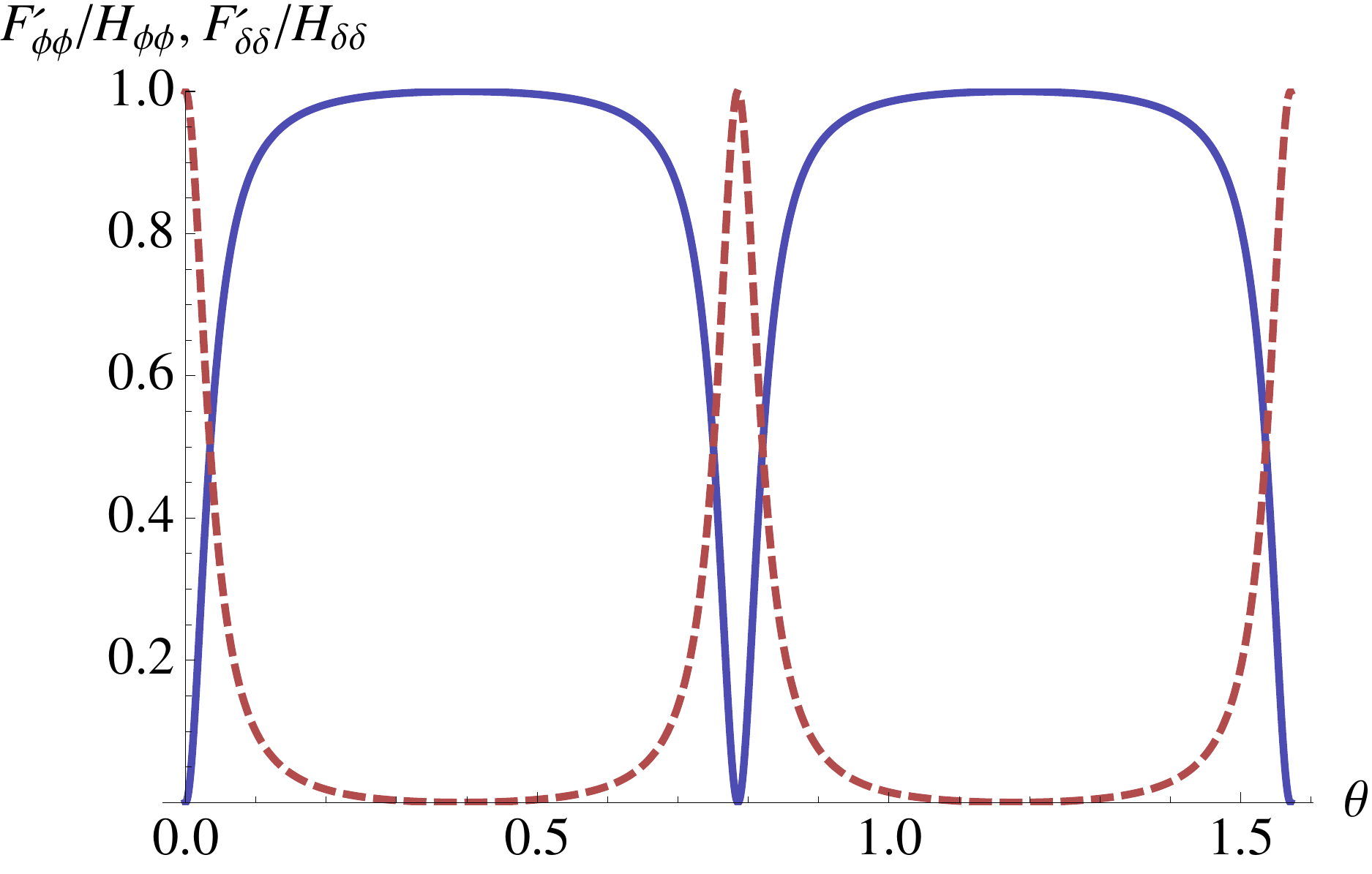} \\
\includegraphics[width=0.9\columnwidth]{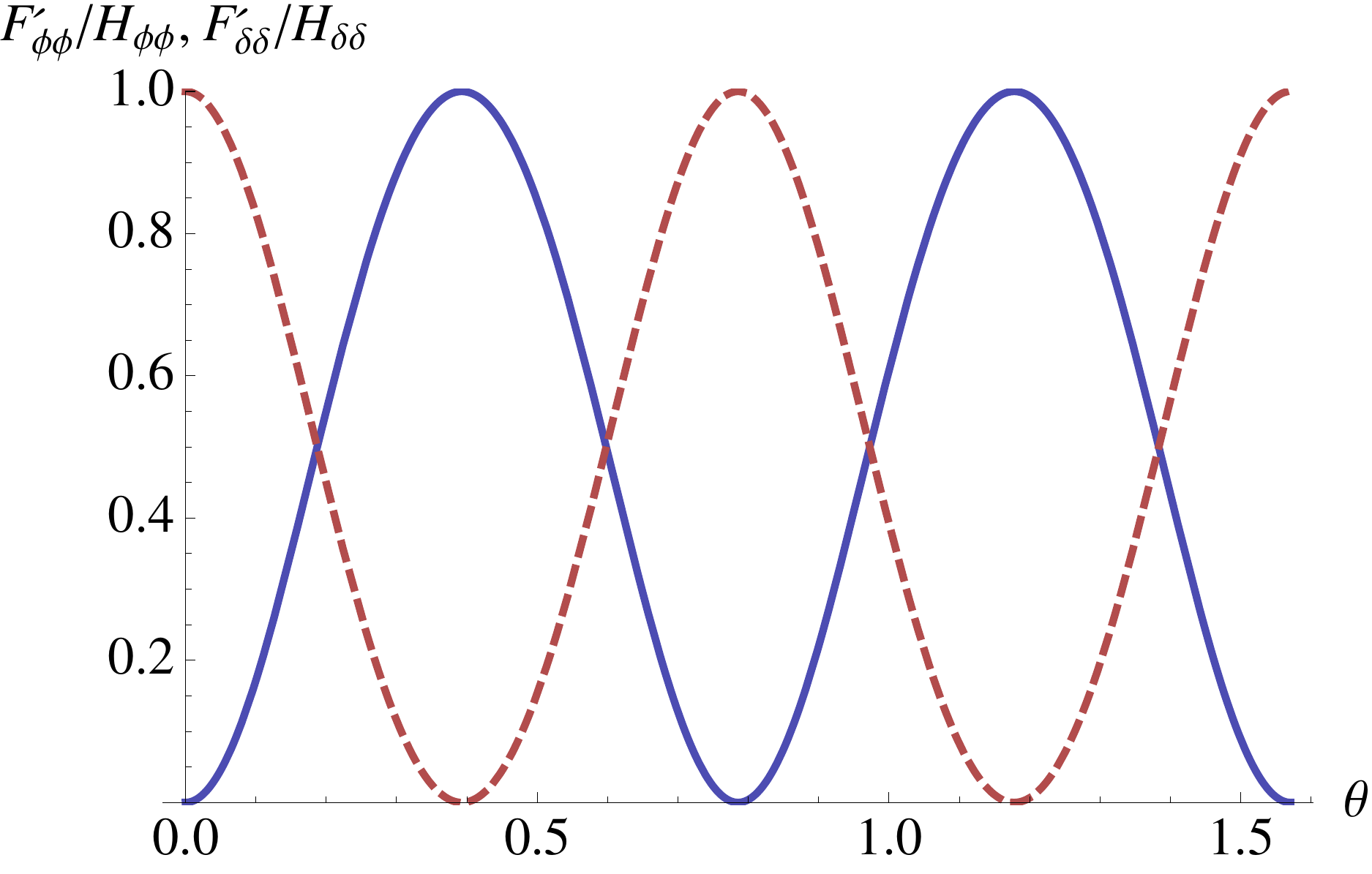} 
\caption{(Color online) Effective Fisher information as a function of the measurement strength $\theta$, rescaled to the corresponding values of the QFI for $\phi=0$ and for different values of $\delta$ (top panel: $\delta=0.1$; bottom panel $\delta=1$). In each plot the blue-solid line corresponds to $F^\prime_{\phi\phi}/H_{\phi\phi}$ while the red-dashed line corresponds to $F^\prime_{\delta,\delta}/H_{\delta,\delta}$. The sum of the two ratios is equal to unit for all values of the measurement strength $\theta$.
\label{fig:2}
}
\end{figure}
\begin{figure}[t]
\includegraphics[width=0.9\columnwidth]{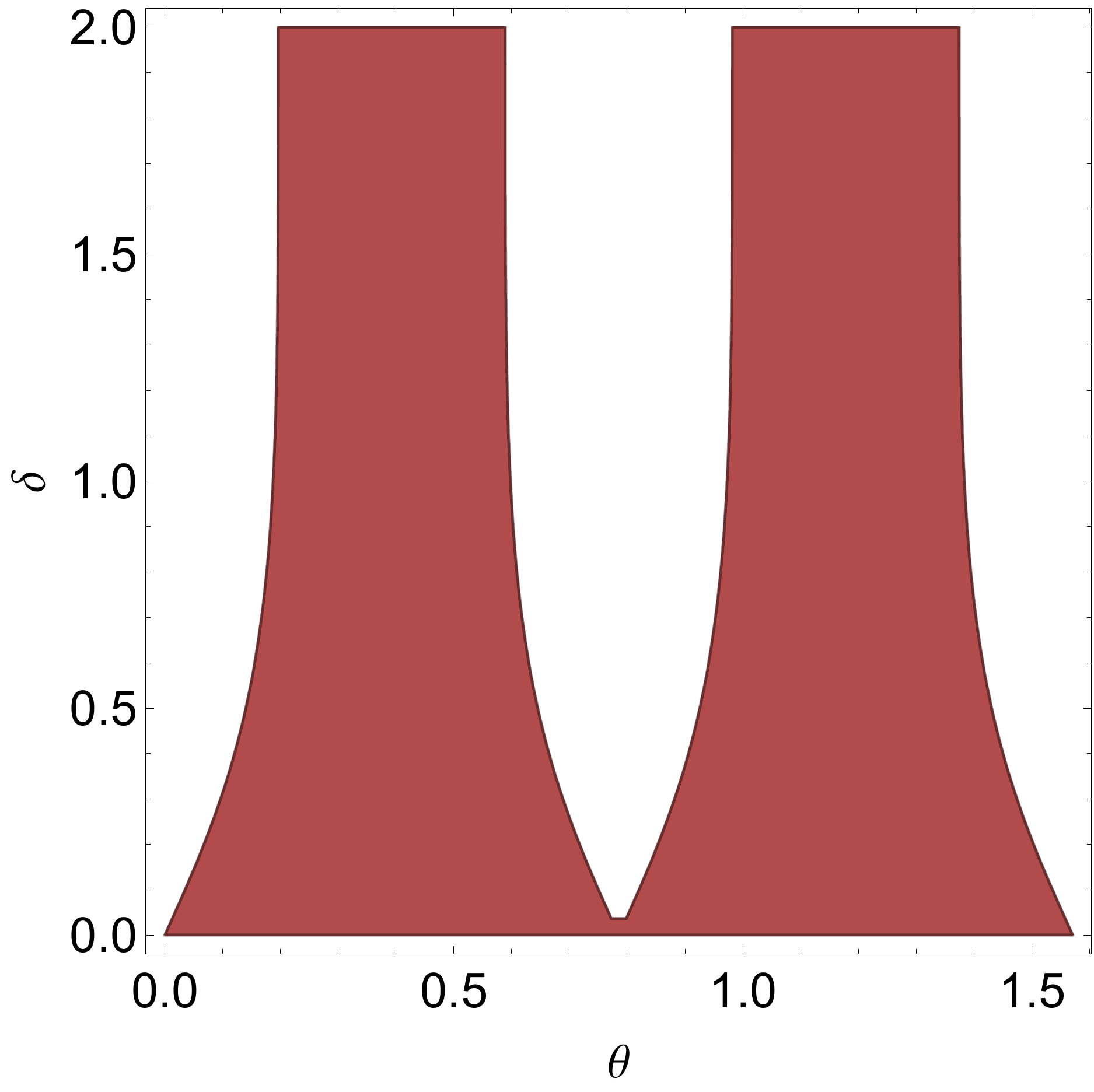} 
\caption{(Color online) The red region correspond to values of the parameters $(\theta, \delta)$ where $F'_{\phi\phi}/H_{\phi\phi}> 1/2$ for $\phi=0$ (in turn the white region corresponds to $F'_{\delta\delta}/H_{\delta\delta}> 1/2$).
\label{fig:3}
}
\end{figure}


Weak measurements offer a different strategy for addressing this joint estimation problem. We can imagine a scheme in which $\sigma_z$ is first measured weakly, and then $\sigma_x$ is measured with a standard projection. The weak measurement provides on average an unbiased estimator~\cite{geoff}, and although it always introduces a disturbance, this can be taken into account in the data analysis by means of some form of calibration, at the cost of a reduction of the attainable precision. We quantify the quality of this measurement by evaluating its Fisher information. To do so, we first introduce weak measurement operators in the form 
\begin{equation}
\label{operatori}
M_\pm=\frac{1}{\sqrt{2}}\left(\cos(\theta/2) \sigma_0\pm\sin(\theta/2)\sigma_z\right), 
\end{equation}
where $\sigma_0$ is the $2{\times}2$ identity matrix. The angle $\theta$ establishes the amount of information extracted in a single run, and it ranges from 0, corresponding to no measurement and no disturbance, to $\pi/2$, corresponding to full projective measurement. These operators allow us to find the final detection probabilities using the modified Born's rule \textcolor{black}{ $p(\pm,\pm){=}\frac{1}{2}\Tr[M_\pm  \varrho  M_\pm(\sigma_0 \pm\sigma_y)]$}, and similarly in the other cases~\cite{geoff,Iinuma}. Our goal here is to relate the Fisher information $\bold F$ to the invasiveness parameters $\theta$ of the weak measurement; the Fisher information can be easily computed from the conditional probabilities $\{ p(+,+),p(+,-),p(-,+), p(-,-)\}$ by using the formula
\begin{align}
F_{\alpha,\beta} &= \sum_{\{j,k\}=\pm} p(j,k) \:(\partial_\alpha \log p(j,k)  )( \partial_\beta \log p(j,k)  ) \nonumber \\ 
&\qquad \textrm{with} \:\: \alpha=\{ \phi, \delta \} \:\: \textrm{and} \:\: \beta=\{\phi,\delta\}.
\end{align}
In particular we obtain.
\begin{widetext}
\begin{align}
&F_{\phi\phi}=\frac{e^{-2\delta^2}}{2}\left(\frac{\cos^2(\theta-\phi)}{1-e^{-2\delta^2}\sin^2(\theta-\phi)}+\frac{\cos^2(\theta+\phi)}{1-e^{-2\delta^2}\sin^2(\theta+\phi)}\right)\\
&F_{\delta\delta}=2\delta^2 e^{-2\delta^2}\left(\frac{\sin^2(\theta-\phi)}{1-e^{-2\delta^2}\sin^2(\theta-\phi)}+\frac{\sin^2(\theta+\phi)}{1-e^{-2\delta^2}\sin^2(\theta+\phi)}\right)\\
&F_{\phi\delta}= F_{\delta\phi}=\frac{\delta e^{-2\delta^2}}{2}\left(\frac{\sin(2\theta-2\phi)}{1-e^{-2\delta^2}\sin^2(\theta-\phi)}-\frac{\sin(2\theta+2\phi)}{1-e^{-2\delta^2}\sin^2(\theta+\phi)}\right)
\end{align}
\end{widetext}

Direct substitution verifies that our weak measurement scheme always saturate, but can not beat the bound \eqref{eq:1}. This fact can be understood by observing that the overall measurement operator, of the form $M_+\cdot(\sigma_0+\sigma_x)\cdot M_+$, corresponds to a projector in the direction $\left(\sin(\theta),0,\cos(\theta)\right)$ of the Bloch sphere, and similar expressions hold in the other cases, originating a symmetry around the $y$ axis; thus, weak measurements are bound by the expression \eqref{eq:1}, and, indeed, saturate it (see the supplemental information of Ref.~\cite{Vidrighin14}). {\color{black} This bound, though, does not give prescriptions as to the behaviour of two-parameter estimation in terms of the strength of the weak measurement, a question which we address in the following.}

The saturation of Eq.\eqref{eq:1} does not ensure that the scheme can actually provide two estimators for $\phi$ and $\delta$, both achieving the \textcolor{black}{optimal bound for the effective Fisher information \eqref{eq:bound2}} . This is because there might exist correlations between these two parameters introduced by the off-diagonal term $F_{\delta\phi}$. 
\textcolor{black}{However one can then directly check that for specific values of the phase, $\phi=k \pi /2$ (with $k \in \mathbbm{Z}$), these correlations are always equal to zero and then the bound (\ref{eq:bound2}) is saturated as well for all values of the other parameter to be estimated $\delta$ and for all values of the measurement strength parameter $\theta$.
By noticing that a phase rotation on the input state is completely equivalent to a rotation of the directions of both the weak measurement and the final strong measurement by the same angle $\phi$, this result shows that the measurement scheme here presented is optimal: one can always saturate the bound by using a two-step adaptive method, which allows to measure the parameters near the optimal working point \cite{Brivio10}. As we show in Figs.~\ref{fig:2} in the case of $\phi=0$, one can tune the measurement strength $\theta$ in order to explore the estimation trade-off and decide the amount of information desired for each of the two parameters. The value of the phase-diffusion can only slightly change the behaviour of the trade-off between the two ratios corresponding to each parameter. In particular, while for small values of $\delta$ the range of values of $\theta$ where one can get more information about the phase $\phi$ is much larger than the one corresponding to more information about the diffusion $\delta$, the trade-off gets more balanced for $\delta \geq 1$. This behaviour is put in evidence in Fig. \ref{fig:3}, where we show the region of values $(\theta,\delta)$ for which the ratio $F^\prime_{\phi\phi}/H_{\phi\phi}$ is larger than $1/2$.}



\section{Experiment} \label{s:exp}
We illustrate the concepts developed in the previous section via a characterisation of a a weak measurement device based on the interferometric scheme introduced by Iinuma and colleagues~\cite{Iinuma}. We perform a weak measurement of the Stokes parameter ${{\cal S}_2=\ket{H}\bra{H}{-}\ket{V}\bra{V}}$ of a light beam, followed by a standard measurement of ${{\cal S}_2=\ket{H{+}V}\bra{H{+}V}{-}\ket{H{-}V}\bra{H{-}V}}$. We are then tackling the problem of estimating the linear polarisation direction, {\it i.e.} a phase shift in the basis of the circular polarisations, together with its associated noise. 

\begin{figure}[b]
\includegraphics[viewport = 80 300 790 540, clip, width=\columnwidth]{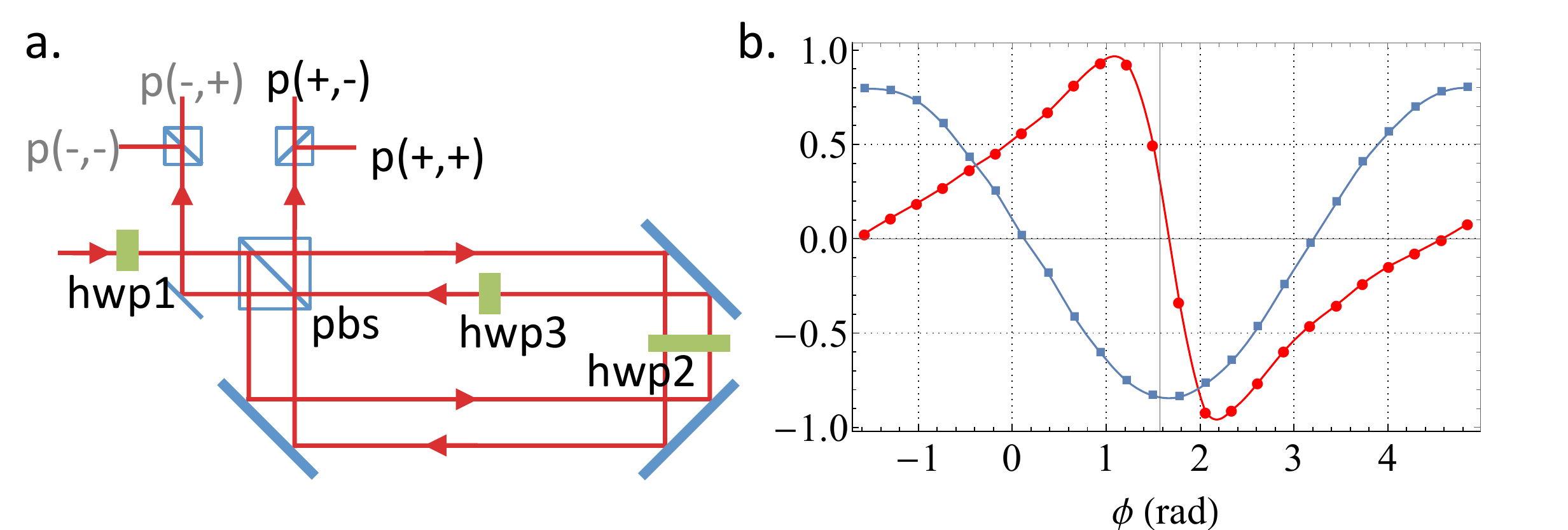}
\caption{(Color online) The measurement device. a) Experimental setup: this is a modification over the original scheme of Iinuma et al.~\cite{Iinuma,Iinuma2}, in which ${\cal S}_1$ is coupled weakly to the path inside the Sagnac interferometer, thus to the two spatial exits. Detection is performed by means of calorimetric power meters; the probabilities $p(-,+)$ and $p(-,-)$ have not been resolved, their sum has been measured instead. b) calibration curves for $p(+,+)+p(+,-)-p(-,+)-p(-,-)$ associated to $\sigma_z$ (blue squares and solid line), and $(p(+,+)-p(+,-))/(p(+,+)+p(+,-))$ (red dots and solid line) associated to $\sigma_x$, as functions of the HWP1 angle $\phi$.}
\label{fig:4}
\end{figure}

A weak measurement necessitates an ancillary metre system to which the observable of interest is coupled. In the original experiment, the path inside a Mach-Zehender interferometer acts as the ancillary system: the beam is first split on a polarising beam splitter (PBS), and then recombined on a beam-splitter, after manipulation of the polarisation~\cite{Iinuma,Iinuma2}. In our arrangement, both the initial and the final beam splitters of the interferometer are PBSs, and that makes it possible to adopt a stable Sagnac configuration, instead of the Mach-Zehender arrangement (Fig. 4a). 

The input polarisation of a cw He:Ne laser is prepared by the first halfwaveplate (HWP1) as an arbitrary linear polarisation $c_H\ket{H}+c_V\ket{V}$. The beam is then divided in two by a PBS, whose outputs constitute the two arms of the Sagnac interferometer. HWP2 is present on both modes, and it sets the measurement strength through its angle $\omega$ with respect the horizontal, while HWP3 is kept with its axis at $-45^\circ$ to ensure that the two spatial modes are coupled differently to the polarisation. With this choice, the effect of the overall device on the polarisation is:
\begin{equation}
\left(
\begin{array}{cc} \cos(2\omega) & 0 \\ 0 & \sin(2\omega)\end{array} 
\right)\cdot 
\left(
\begin{array}{c} c_H \\ c_V \end{array} 
\right)
\end{equation} 
for output 1, and 
\begin{equation}
\sigma_x \cdot \left(
\begin{array}{cc} \sin(2\omega) & 0 \\ 0 & \cos(2\omega)\end{array} 
\right)\cdot 
\left(
\begin{array}{c} c_H \\ c_V \end{array} 
\right)
\end{equation}
for output 2. Therefore we obtain an implementation of the measurement operators \eqref{operatori}, up to a unitary; the orientation of the axis of HWP2 is linked to the setting of the measurement by $\theta{=}\frac{\pi}{2}-4\omega$, which is set in our experiment to $\theta_0{=}58^\circ$, and approaches the strong regime. Due to technical limitations, we only consider a three-outcome measurement, without resolving $\sigma_x$ on one of the outputs. This results in a slight decrease of the available Fisher information, but, notably, the correlations between the two parameters, which are our main interest for this investigation, remain unaffected.

\begin{figure}[t]
\includegraphics[width=\columnwidth]{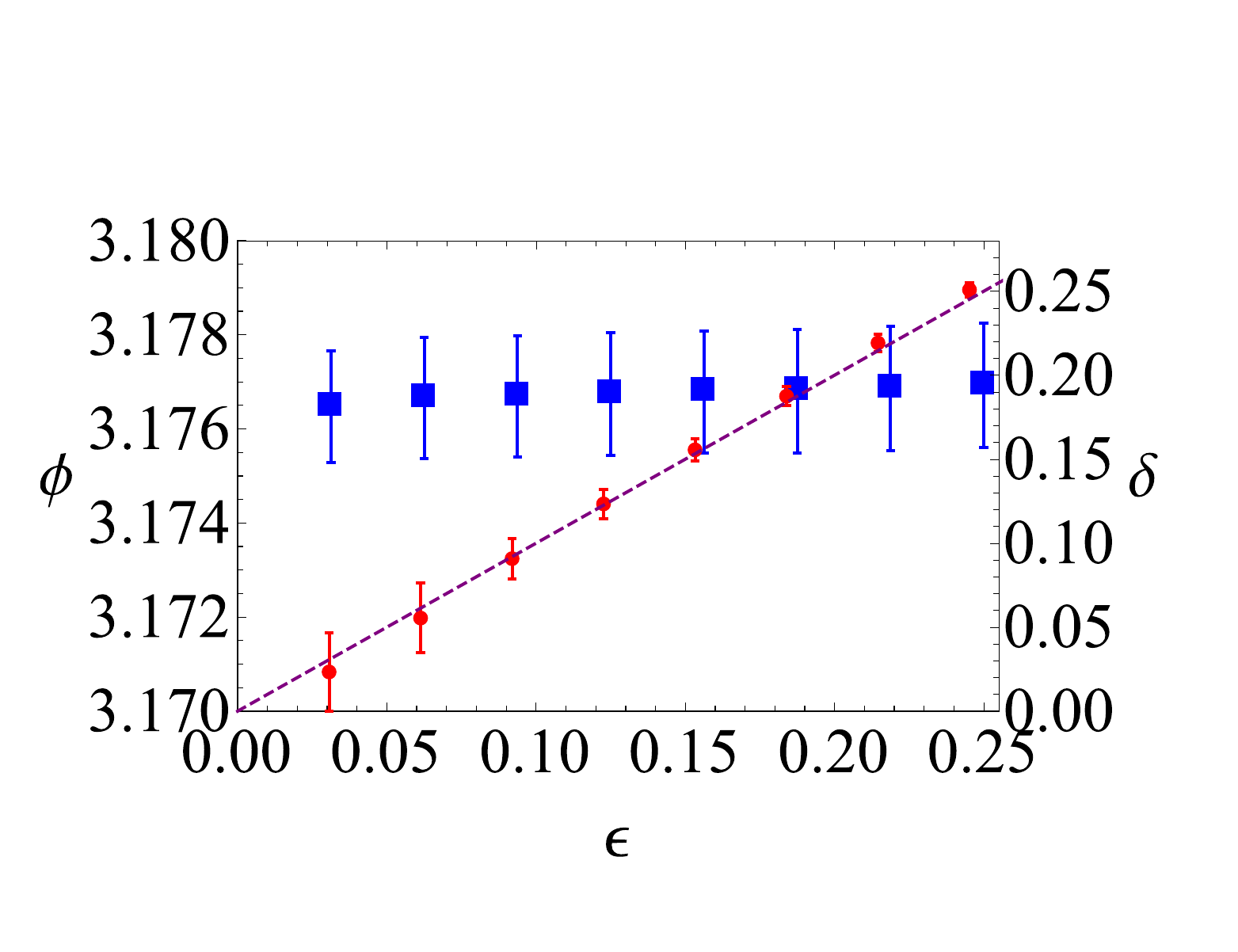}
\caption{(Color online) Experimental multiparameter estimation. The mixed state \eqref{stato} is simulated by adding the data corresponding to pure state with $\theta\simeq{3.18}$, and $\theta\simeq{3.18-\pi}$, with weights $\frac{1+e^{-\delta_0^2}}{2}$, and $\frac{1-e^{-\delta_0^2}}{2}$, respectively. For each point the expected value of the dephasing is set to a value $\delta_0$, and the parameters are estimated by a minimal residual based on the curves in Fig.(2), implemented by maximum likelihood algorithm; this delivers the values $\phi$ and $\delta$. The dashed curve shows the expected behaviour $\delta{=}\delta_0$. The uncertainties are estimated with a Monte Carlo procedure: we simulate 10000 repetitions of the experiment in which the intensities vary within the experimental uncertainty.}
\label{fig:5}
\end{figure}

\begin{figure*}[h]
\subfigure[]{\includegraphics[width=2\columnwidth, viewport = 50 130 1150 400, clip]{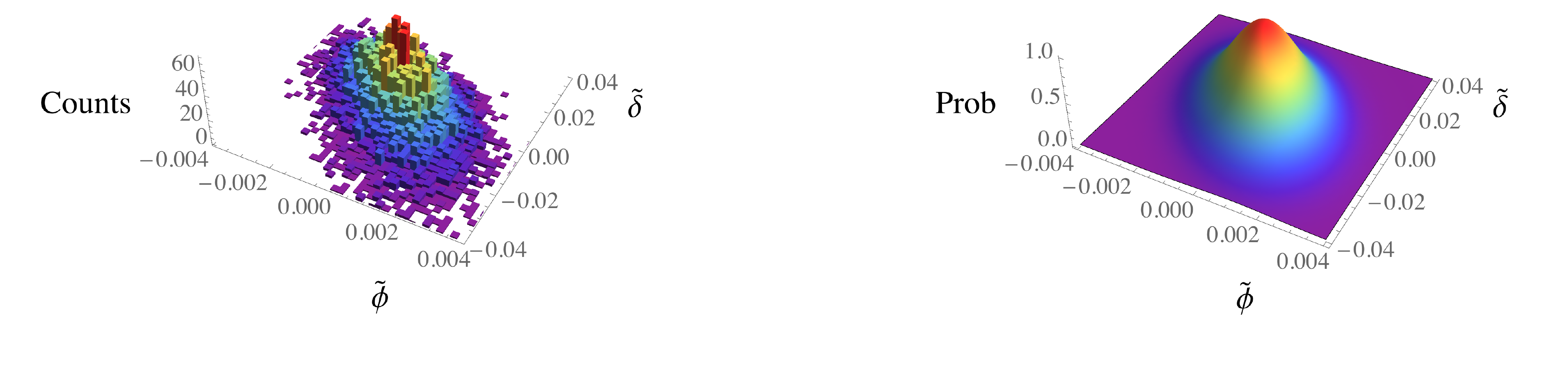}}
\subfigure[]{\includegraphics[width=2\columnwidth, viewport = 50 130 1150 400, clip]{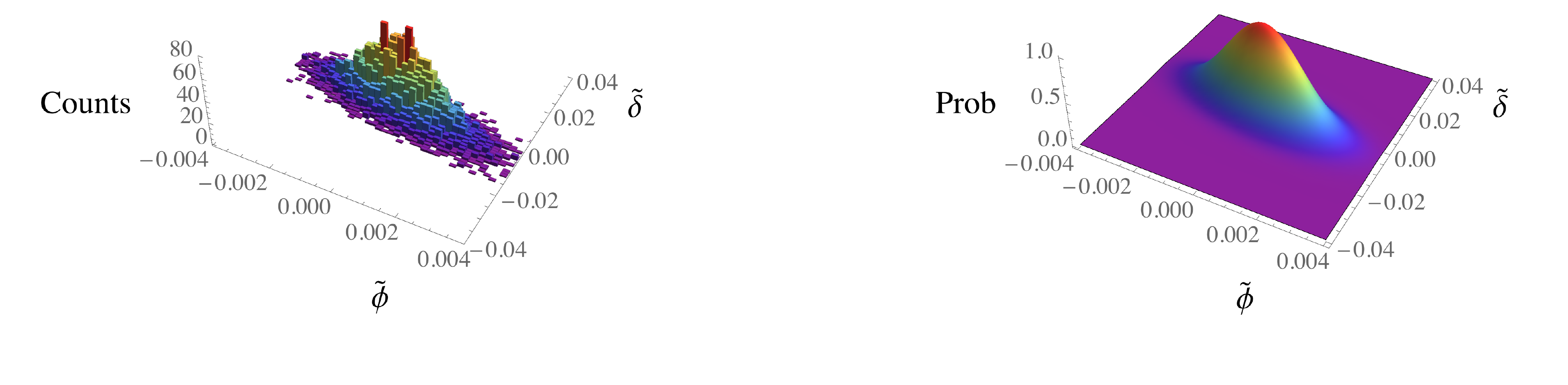}}
\caption{(Color online) Histograms of the Monte Carlo realisations for (a) $\delta_0{=}0.094$, and (b) $\delta_0{=}0.25$, compared with the expected covariances, as a function of the deviations $\tilde\phi$ and $\tilde\delta$ from the mean values. The covariance is calculated for $M'{=}4\cdot10^5$; it serves as a comparison of the correlations between the two estimates, and does not correspond to the Cram\'er-Rao bound, since $M'$ is lower than the average photon number as an effect of extra-noise. We observe that qualitatively the variation in the correlations follows the prediction for an ideal weak measurement device. It is worth remarking that when working at low $\delta_0 \lesssim 0.05$ the optimisation procedure that extracts $\{\phi,\delta\}$ sometimes gets trapped in a local minimum in correspondence of $\delta=0$.}
\label{fig:6}
\end{figure*}

As a first step, we perform a calibration of our device, by injecting pure states with known phases set by the angle of HWP1, and register the outcome probability. We then obtain calibration curves such as those in Fig.~4b, using an interpolation~\cite{nota}; arbitrary pairs of  phase and dephasing can then be recovered by finding the values of $\phi$ and $\delta$ which are more compatible with these curves with a minimal residual approach. Following the technique in~\cite{qcontrol}, in our experiment we have simulated the state \eqref{stato} by injecting two pure states and adding the output signals with proper weighting.  Figure \ref{fig:5} reports the experimental values $\phi$ and $\delta$ obtained when varying the expected dephasing $\delta_0$. As expected, the estimated phase does not change with the amplitude of the dephasing, while we observe a small bias in the estimation of $\delta$, due to the fact that the data sets have not the same intensity.

In a multiparameter estimation experiment, an important aspect is the manifestation of correlations between the estimated parameters. {\color{black} Since the main source of uncertainty in our experiment is the discretisation imposed by the intensity meter, a repetition of the experiment would not lead to an insight nor on the errors on each parameter, nor on their correlations. In order to obtain a visualisation of such uncertainties we have employed a Monte Carlo routine which simulates repetitions of the estimation experiment, by varying the intensity values within the range of the experimental uncertainties. This constitutes a simple and direct way to inspect the experimental covariance of the measured parameters. We then compare these histograms with the expected covariance $\Sigma=F/M'$ for an ideal instrument. We emphasise that, due to the origin of the uncertainties in our measurement, the parameter $M'$ should not be interpreted directly as an effective number of experiments, but as a scaling parameter to make the two figures comparable~\cite{note}. As expected, the shape of the correlations depends on the working value of $\delta_0$, and the trend is followed even in the presence of imperfections.}

\section{conclusions}\label{s:conclusions}
We have discussed the performance of weak measurements as a scheme for the joint estimation of a phase parameter and the amplitude of contextual dephasing. We have illustrated how these measurements obey the same bound as all POVMs, and related the information obtained on the two parameters to the measurement setting. An experimental investigation has shown how these considerations remain valid qualitatively even in presence of experimental imperfections. Our results highlight how these measurements represent a practical scheme for such multiparameter scenario, but present no intrinsic advantage.

\section*{Acknowledgements}
We thank Animesh Datta for discussion and comments, Paolo Aloe for technical assistance, Sandro Giacomini, Paolo Mataloni, Francesca Paolucci, Alfonso Russo and Maria Antonietta Ricci for the loan of optical equipment. M.B. has been supported by a Rita Levi-Montalcini fellowship of MIUR, and by encouragement from Alessandro Alberucci. M.D.V. is supported by the EPSRC {\it via} the Controlled Quantum Dynamics CDT. M.G.G. acknowledges support from EPSRC through grant EP/K026267/1.

\appendix
\section{Multiparameter quantum estimation theory} \label{s:appendixLQE}
\textcolor{black}{
Let us consider a physical system parametrized by a set of $n$ parameters ${\boldsymbol \lambda} = \{ \lambda_1, \dots , \lambda_n \}$. An estimator $\hat{\boldsymbol \lambda}=\{ \hat{\lambda}_1, \dots, \hat{\lambda}_n \}$ is defined as a map from a set of $M$ measurement outcomes $X = \{x_1, \dots, x_M \}$ to the possible values of the parameters $\boldsymbol{\lambda}$. We consider unbiased estimators, {\em i.e.} estimators such that, for a sufficiently large number of measurements $M$, satisfy the property $\mathbbm{E}[\hat{\boldsymbol \lambda}] = {\boldsymbol \lambda}$  ($\mathbbm{E}[\cdot]$ denoting the average over all the possible measurement results $X$). The performances of unbiased estimators is quantified in terms of the covariance matrix ${\bf \Sigma}$, whose elements are defined as 
\begin{equation}
\Sigma_{ij} = \mathbbm{E}[ (\hat{\lambda}_i - \lambda_i ) (\hat{\lambda}_j - \lambda_j) ] \:,
\end{equation}
such that, for unbiased estimators, the diagonal elements correspond to the variances and off-diagonal elements to covariances. The covariance matrix is lower bounded according to the Cram\'er-Rao bound as
\begin{equation}
{\bf \Sigma} \geq \frac{{\bf F}^{-1}}{M} \label{eq:CRBClassical}
\end{equation}
where ${\bf F}$ denotes the Fisher information matrix, with components
\begin{equation}
F_{ij} = \int dx \: p(x | {\boldsymbol \lambda} ) ( \partial_{\lambda_i} \log p(x | {\boldsymbol \lambda} ) )  (\partial_{\lambda_j} \log p(x | {\boldsymbol \lambda} )  ) \:. \label{eq:classicalF}
\end{equation}
In quantum mechanics the conditional probability entering in Eq. (\ref{eq:classicalF}) can be recasted as $p(x|{\boldsymbol \lambda}) = \hbox{Tr}[\varrho_{\boldsymbol \lambda} \Pi_x]$, where $\varrho_{\boldsymbol \lambda}$ is the quantum state parametrized in terms of $\boldsymbol \lambda$, and $\{\Pi_x\}$ is the POVM corresponding to the measurement performed. By defining the symmetric logarithmic derivative (SLD) operators by the equation
$L_k \varrho_{\boldsymbol \lambda} + \varrho_{\boldsymbol \lambda} L_k = 2 \partial_{\lambda_k} \varrho_{\boldsymbol \lambda}$, one can introduce the quantum Fisher information (QFI) matrix ${\bf H}$, with elements
\begin{equation}
H_{ij} = {\rm Re} \left( \hbox{Tr}[ \varrho_{\boldsymbol \lambda} L_i L_j ] \right) \:.
\end{equation}
The QFI matrix provides a more general bound for the covariance matrix, holding for any possible POVM, {\em i.e.}
\begin{equation}
{\bf \Sigma} \geq \frac{{\bf H}^{-1}}{M} \:. \label{eq:CRBQuantuml}
\end{equation}
While the classical bound (\ref{eq:CRBClassical}) can always be saturated, for example with a maximum likelihood estimator, the quantum bound is proven to be always achievable only for a single parameter, with the optimal measurement corresponding to the eigenstates of the SLD operator. In the multi-parameter case, the optimal measurements for each parameter (still corresponding to the eigenstates of the different SLD operators) may not commute and as a consequence the bound cannot in general be saturated. Moreover, different bounds can be obtained by considering different derivation operators~\cite{GenoKim}; in our case, the SLD provides the most informative bound.
}


\begin{thebibliography}{99}
\bibitem{nota} We have used the function Interpolation in the Mathematica software package, with an order 3 interpolation between points.
\bibitem{note} Since the uncertainties are associated mostly to the discretisation of the readings, a flat distribution would represent the data more accurately. For the purpose of clarity, we have  adopted a Gaussian distribution with an equivalent variance instead. This changes the shape of the histograms without affecting their characteristic widths, and offers a more direct comparison
with the expected covariance.
\end{thebibliography}
\end{document}